\documentclass[fleqn,twoside]{article}
\usepackage{espcrc2,epsfig}
\usepackage{graphicx}
\usepackage[figuresright]{rotating}
\newcommand{\be}{\begin{equation}}
\newcommand{\ee}{\end{equation}}
\newcommand{\vev}[1]{\left\langle #1 \right\rangle}
\newcommand{\tr}{{\rm Tr}}
\newcommand{\m}{\hphantom{$-$}}

\title{Lattice supersymmetric Ward identities}

\author{Federico Farchioni\address[DESY]{Deutsches Elektronen-Synchrotron, DESY,
        D-22603 Hamburg, Germany}\thanks{Talk given by Federico Farchioni.
        Address after October 1st: Institut f\"ur Theoretische Physik, Universit\"at M\"unster,
        Wilhelm-Klemm-Str. 9, D-48149 M\"unster, Germany.}, 
        Alessandra Feo\address[UM]{Institut f\"ur Theoretische Physik,
        Universit\"at M\"unster,
       \\
        Wilhelm-Klemm-Str. 9, D-48149 M\"unster, Germany},
        Tobias Galla\address[UO]{Department of Physics,
        University of Oxford, 1 Keble Road, Oxford OX1 3NP, UK},
        Claus Gebert\addressmark[DESY], Robert Kirchner\addressmark[DESY]
        \thanks{Address after October 1st: Universidad Aut\`onoma de Madrid, 
        Cantoblanco, Madrid 28049, Spain.},
       \\
        Istv\'an Montvay\addressmark[DESY], Gernot M\"unster\addressmark[UM],
        Anastassios Vladikas\address[URTV]{INFN, Sezione di Roma 2, 
        Universit\'a di Roma ``Tor Vergata'', I-00133 Rome, Italy},
       \\
        [0.5em]
        DESY-M\"unster-Roma Collaboration \\[0.5em] }

\begin{document}

\begin{abstract}

 SUSY Ward identities for the N=1 SU(2) 
 SUSY Yang-Mills theory are studied on the lattice
 in a non-perturbative numerical approach. 
 As a result a determination of the subtracted gluino mass is obtained.

\end{abstract}

\maketitle

\section{Introduction}

The formulation of SUSY gauge theories on the lattice 
is problematic since the discretization breaks the Poincar\'e 
invariance, a sector of the superalgebra. 
In the Wilson approach the suppression of unphysical states
in the fermionic sector
is obtained by the introduction of an
extra-term (Wilson term) which explicitly breaks SUSY.
The restoration of SUSY in the continuum limit 
can be verified by considering the related
lattice Ward identities (SUSY WIs)~\cite{WI}.

We focus on  the N=1 SU(2) SUSY Yang-Mills theory (SYM)
(see also~\cite{past} and references therein).
This is the SUSY version of quantum gluodynamics 
where gluons are accompanied by 
fermionic partners (gluinos) in the same (adjoint) 
representation of the color group.
As a consequence of the explicit breaking of the symmetry, 
the SUSY WIs assume in the lattice theory a peculiar form. 
We restrict the analysis to the on-shell regime~\cite{DoetAl}.
A subtracted gluino mass $m_S$ appears; in addition, the SUSY current
$S_{\mu}(x)$ gets a multiplicative factor
$Z_S$ and a new mixing term $Z_T\partial_\mu T_{\mu}(x)$
is added to the nominal WIs of the continuum.

In this contribution we present the non-perturbative
determination of the quantities $m_SZ_S^{-1}$ and $Z_TZ_S^{-1}$
from the numerical analysis of the SUSY WIs.
Preliminary results were presented in~\cite{proc}.
More details, including related theoretical issues, will be presented in a forthcoming 
publication. This study is also complemented by a perturbative
computation~\cite{pert}.

The numerical computations were performed on the
CRAY-T3E computers at John von Neumann Institute for Computing (NIC),
J\"ulich. We thank NIC and the staff at ZAM for their kind support.

\section{Method}

\begin{figure}[t]
\includegraphics*[width=13pc,angle=-90]{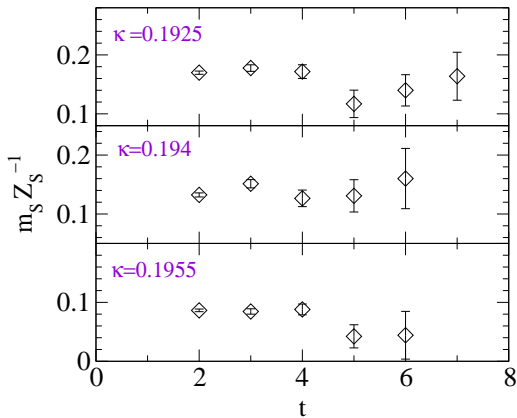}
\vspace*{-5mm}
\caption{$m_SZ_S^{-1}$ as a function of $t$
with insertion $\chi^{(sp)}(x)$ (point-split currents).}
\label{fig:m_time} 
\vspace*{-5mm}
\end{figure}

We consider the zero momentum lattice SUSY WI with insertion ${\cal O}(y)$
\begin{eqnarray}\label{renormward}\nonumber
&\!\!\!\!\!\!Z_S\!\!\!\!\!\!&\sum_{\vec{x}}\!\!\vev{\left(\nabla_0 S^{l}_0(x)\right){\cal O}(y)}\!+\!
Z_T\!\!\sum_{\vec{x}}\!\!\vev{\left(\nabla_0 T^{l}_0(x)\right) {\cal O}(y)} \\
&\!\!\!\!=\!\!\!\!& m_S\sum_{\vec{x}}\!\!\vev{\chi^{l}(x) {\cal O}(y)}\,+\,O(a)\ .
\end{eqnarray}
This WI is valid in the on-shell regime where $x\!\neq\! y$
and for gauge-invariant operators ${\cal O}(x)$
(see~\cite{pert} for the general case).
We explain briefly the meaning of the various quantities in~(\ref{renormward}).
$S^{l}_\mu(x)$, $T^{l}_\mu(x)$ and $\chi^{l}(x)$ (sink operators)
are lattice forms of the SUSY current 
$S_\mu(x)=-\sigma_{\rho\sigma}\gamma_\mu\tr(F_{\rho\sigma}(x) \lambda(x))$, 
the mixing current $T_\mu(x)=2\gamma_\nu\tr(F_{\mu\nu}(x) \lambda(x))$ 
and the soft breaking operator $\chi(x)=\sigma_{\rho\sigma}\tr(F_{\rho\sigma}(x) \lambda(x))$,
respectively.
The trace is taken on color indices and $\lambda(x)$ is the 
adjoint Majorana field of the gluino.
We consider~\cite{proc} a local and a point-split definition of the currents.
The field tensor $F_{\mu\nu}(x)$ is replaced by a clover-symmetric
lattice field tensor. 
The quantities $Z_S$ and $Z_T$ are renormalizations coming
from the lattice SUSY breaking, $m_S$ is the gluino mass
shifted by additive renormalization. 
The condition $m_S=0$ is supposed to correspond, 
in the continuum limit, to the physical situation where the gluino 
is massless and SUSY is restored.

In our analysis we consider  the lowest dimensional insertion operators ${\cal O}(x)$
($d=7/2$). One has essentially two choices~\cite{proc}. One is
the time-slice operator obtained from $\chi^{l}(x)$
by discarding time-like plaquettes,
${\cal O}^{(1)}(x)=\chi^{(sp)}(x)$; another possibility is
${\cal O}^{(2)}(x)=T_0^{(loc)}(x)$, extended in the time-direction.
We smear the insertion operators by combined APE and Jacobi 
smearing on the gluon and gluino fields respectively.
Smearing significantly improves the signal for $\chi^{(sp)}(x)$
but not for $T_0^{(loc)}(x)$.
This is presumably because the latter contains temporal links,
for which a multi-hit procedure is more appropriate than smearing.
Such a procedure is however computationally too expensive in our setup 
with dynamical fermions.

For a given insertion ${\cal O}(x)$ 
the WI~(\ref{renormward}) results in two independent equations
when composing the spins of sink and insertion operators.
The solution of the $2\!\times\! 2$ linear system allows 
the non perturbative determination of $m_SZ_S^{-1}$ and $Z_TZ_S^{-1}$
for each time-separation $t\!=\!x_0-y_0$.
See Fig.~\ref{fig:m_time} for an example.
Alternatively we solve the overdetermined linear system
for several time-separations $(t_{min},\cdots, L_t/2$);
the values of  $m_SZ_S^{-1}$ and $Z_TZ_S^{-1}$ are obtained in this way
through a least-square fit.

\section{Results}

\begin{table*}[htb]
\caption{Summary of results.}
\label{tab}
\renewcommand{\tabcolsep}{1pc} 
\begin{tabular}{@{}lcllll}
\hline
& & \multicolumn{2}{c}{point split currents} &\multicolumn{2}{c}{local currents} \\ 
\hline
\multicolumn{1}{c}{$\kappa$} & operator & \multicolumn{1}{c}{$m_SZ_S^{-1}$}  
& \multicolumn{1}{c}{$Z_TZ_S^{-1}$} & \multicolumn{1}{c}{$m_SZ_S^{-1}$}  & \multicolumn{1}{c}{$Z_TZ_S^{-1}$}\\
\hline
0.1925 & $\chi^{(sp)}$      & 0.176(5)   & $-$0.015(19) & 0.166(6)  & 0.183(14)     \\    
0.1925 & $\chi^{(sp)}$ (*)  & 0.182(6)   & $-$0.044(16) &  0.173(6)  & 0.176(14)     \\    
0.1925 & $\chi^{(sp)}$ (**) & 0.1969(47) & $-$0.058(14) &  0.1821(47)& 0.146(11)     \\   
0.1925 & $T^{(loc)}_0$      & 0.132(16)  & \m 0.11(7)   &  0.144(18) & 0.29(6)     \\    
\hline
0.194  & $\chi^{(sp)}$      & 0.148(6)   & $-$0.038(19) &  0.124(6)  & 0.202(15)     \\    
0.194  & $T^{(loc)}_0$      & 0.095(27)  & \m 0.11(13)  &  0.076(30) & 0.27(9)      \\  
\hline
0.1955 & $\chi^{(sp)}$      & 0.0839(4)  & $-$0.051(13) &  0.0532(40) & 0.179(10)     \\ 
\hline
\end{tabular} \\[2pt]
\parbox{11cm}{* \footnotesize With plaquette field tensor.}\\
\parbox{11cm}{** \footnotesize With plaquette field tensor and different smearing.}\\
\end{table*}

Configurations were generated on a $12^3\times24$ lattice at 
$\beta=2.3$ by means of the two-step multi-bosonic algorithm (TSMB).
See~\cite{past} and references therein for more details on the
algorithm. 
See also~\cite{qcd3f} for an application to QCD with three dynamical quark flavors. 
The configurations at $\kappa\!=\!0.1925$ were produced in~\cite{past}.
Results concerning $\kappa\!=\!0.1925$ and 0.194 were already presented in~\cite{proc}.
We add here more statistics at $\kappa\!=\!0.194$ and a new simulation point,
$\kappa\!=\!0.1955$. 
The algorithmic setup was optimized in order to reduce autocorrelations 
for light gluinos.

In Table~\ref{tab} we report the complete results for
the global fit over a range of time-separations.
The smallest time-separation included in the fit $t_{min}$
was chosen such that contact terms in the correlations are absent;
this means $t_{min}=3$ for insertion $\chi^{(sp)}(x)$ and 
$t_{min}=4$ for $T^{(loc)}_0(x)$.
Discretization effects can be evaluated by comparing 
determinations from different insertions,
see data for $\kappa=0.1925$ and $\kappa=0.194$. For $\kappa=0.1925$
we also report results for different definitions of $\chi^{(sp)}(x)$,
namely for the simple-plaquette definition of the
field tensor and for different smearing parameters.
The deviation ranges between 
$20\%$ and $40\%$ for $m_SZ_S^{-1}$. Data from $T^{(loc)}_0(x)$ are however 
subject to large statistical fluctuations and 
thus $O(a)$ effects cannot be reliably estimated.

\begin{figure}[t]
\includegraphics*[width=15pc]{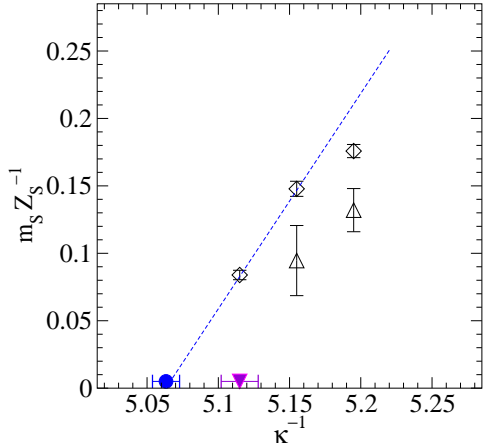}
\vspace*{-5mm}
\caption{$m_SZ_S^{-1}$ as a function of $\kappa^{-1}$
with insertion $\chi^{(sp)}(x)$ (diamonds) 
and  $T_0^{(loc)}(x)$ (triangles) and point-split currents.
The filled circle is the result of the extrapolation,
the filled triangle is the determination of $\kappa_c$
of~\cite{KietAl}.}
\vspace*{-5mm}
\label{fig} 
\end{figure}

In Fig.~\ref{fig} we report the determination
of $m_SZ_S^{-1}$ as a function of the inverse hopping parameter. 
The expectation is that $m_SZ_S^{-1}$ vanishes linearly when 
$\kappa\rightarrow\kappa_c$. 
We see a clear decrease when $\kappa$ is increased towards $\kappa_c$.
An extrapolation using
data from insertion $\chi^{(sp)}(x)$ gives as a result:
$\kappa_c=0.19750(38)$ for the point-split currents
and $\kappa_c=0.19647(27)$ for the local ones.
The result can be compared with the estimate
$\kappa_c=0.1955(5)$ from the study of the first order phase 
transition~\cite{KietAl}.
An analogous analysis for the quantity  $Z_TZ_S^{-1}$
(fitting to a constant, in this case)
gives $Z_TZ_S^{-1}=-0.039(7)$ for
the point-split currents and
$Z_TZ_S^{-1}=0.185(7)$ for the local ones.

Our results demonstrate the feasibility of implementing lattice SUSY WIs
in order to verify supersymmetry restoration in a non-perturbative
framework.

\end{document}